\documentclass[fleqn,twoside]{article}
\usepackage{espcrc2}
\usepackage{epsfig}

\usepackage{graphicx}
\usepackage[figuresright]{rotating}

\def\beq{\begin{equation}}
\def\eeq{\end{equation}}
\def\bea{\begin{eqnarray}}
\def\eea{\end{eqnarray}}
\def\bq{\begin{quote}}
\def\eq{\end{quote}}

\newlength{\bredde}
\def\slash#1{\settowidth{\bredde}{$#1$}\ifmmode\,\raisebox{.15ex}{/}
\hspace*{-\bredde} #1\else$\,\raisebox{.15ex}{/}\hspace*{-\bredde} #1$\fi}

\def\gappeq{\mathrel{\rlap {\raise.5ex\hbox{$>$}}
{\lower.5ex\hbox{$\sim$}}}}

\def\lappeq{\mathrel{\rlap{\raise.5ex\hbox{$<$}}
{\lower.5ex\hbox{$\sim$}}}}

\def\Toprel#1\over#2{\mathrel{\mathop{#2}\limits^{#1}}}

\newcommand{\AmS}{{\protect\the\textfont2
  A\kern-.1667em\lower.5ex\hbox{M}\kern-.125emS}}

\title{\vspace{-4.0cm}
       \rightline{\normalsize CERN-TH/2001-286}
        \rightline{\normalsize DESY 01-163}
	 \rightline{\normalsize IFIC/01-57, FTUV/011019}
       \vspace{2.0cm}
Finite-volume meson propagators in quenched chiral
perturbation theory}

\author{P.H. Damgaard \address[cern]{Theory Division,
CERN, 1211 Geneva 23, Switzerland}\thanks{On leave from the Niels Bohr
Institute, Blegdamsvej 17, DK-2100 Copenhagen, Denmark.}, M.C.
Diamantini\addressmark[cern]\thanks{Swiss National Science Foundation
fellow.},
P.~Hern\'andez\addressmark[cern]\thanks{On leave from Dpto. F\'{\i}sica
Te\'orica, Univ. Valencia.}, K. Jansen\address[]{NIC/DESY Zeuthen,
Platanenallee 6, D-15738 Zeuthen, Germany} }

\begin{document}

\begin{abstract}
We compute meson propagators in
finite-volume quenched chiral perturbation theory.
\vspace{1pc}
\end{abstract}

\maketitle


Recent progress in the formulation of chiral fermions on the lattice makes
it possible to approach the regime of the physical light quark masses.
However,
lattice volumes $V=L^4$ needed to ensure negligible finite size
effects are prohibitively large for such light quarks. Only when $L$
is very large compared to the Compton wavelength of the lightest
particles of the theory (i.e. the pions) are the volume effects exponentially
suppressed. Fortunately chiral perturbation theory ($\chi$PT) can predict
the volume effects that occur in the regime $m_\pi L \leq 1$.
In this way the finite volume becomes a distinct advantage, and
a finite-size scaling analysis is actually a very useful tool to extract
the physical low energy constants. This has been demonstrated
in simpler models \cite{on}.

The $\chi$PT predictions in full QCD for the finite size scaling of
quantities such as the chiral
 condensate or the propagators of scalar and vector densities have been
computed in \cite{gl,hl}. Unfortunately, lattice simulations
in the regime of very light quark masses are presently only possible in the
quenched approximation, and the predictions of chiral
perturbation theory are quite drastically modified in this approximation
\cite{BGS}.
We present here some results for the
propagators of the scalar and pseudoscalar densities at finite volume
in quenched $\chi$PT (Q$\chi$PT). Full details will be presented 
in a forthcoming publication.


The low-energy limit of QCD with light quarks is a chiral Lagrangian
of the Goldstone bosons resulting from spontaneous chiral symmetry breaking.
In an infinite volume, the effective Lagrangian is approximated by an
expansion in powers of the pion momentum $p$ and mass
$m_\pi$. This is the standard {\it p-expansion}.
In a finite volume the $p$-expansion is still good if the Compton wavelength
of the particle is much
smaller than the size of the box ( $L\gg {1\over m_\pi}$ ). In the
opposite limit,
$L\ll {1\over m_\pi}$, the $p$-expansion breaks down due to propagation
of pions with  zero momenta \cite{gl}.
A convenient expansion for this
regime is the so-called {\it $\epsilon$-expansion} in which $m_\pi \sim p^2
\sim \epsilon^2$ \cite{gl}. By the Gell-Mann--Oakes--Renner relation
this corresponds to keeping $\mu_i\equiv m_i\Sigma V$ of order unity while
$V$ is taken large.
The zero mode of the pion can be isolated by
factorizing  $U(x) = U_0 \exp i \sqrt{2}\xi(x)/F_{\pi}$ into the  constant
collective field $U_0$ and  the pion fluctuations $\xi (x)$.
The difficulty then comes from the fact that
the integral over $U_0$ needs to be done exactly, while ordinary
$\chi$PT applies to the non-zero mode integration.
At leading order one obtains the suitably normalized partition function
\bea
{\cal Z} = \int_{SU(N_f)} dU_0 \exp \left[ {\Sigma \over 2} V {\rm Tr }
M\left( U_0 + U_0^+\right) \right].\nonumber
\eea
where $M$ is the quark mass matrix.
It is interesting to consider averages in sectors of fixed
topology $\nu$ as well \cite{ls}. To the same order
\bea
{\cal Z}_\nu \!=\! \int_{U(N_f)} \hspace{-.5cm} dU_0 \left({\rm det}
U_0\right)^\nu \exp\! \left[
{\Sigma \over 2} V {\rm Tr } M\left( U_0 +
U_0^+\right )\! \right]. \nonumber
\eea

The $\epsilon$-expansion can be worked out also in the case of
Q$\chi$PT. We have considered two methods to take the quenched limit. The
first is the supersymmetric method \cite{BGS}, suitably refined to be valid
at the non-perturbative level \cite{OTV}. Here the fermionic determinant
of $N_v$ valence quarks is cancelled
by adding to the theory $N_v$ flavors of ghost bosonic quarks. Assuming
a supergroup generalization of the
chiral symmetry breaking pattern of QCD, the field $U(x)$ becomes an
element of a graded group with both Goldstone bosons and Goldstone fermions.
The integration over these fields is taken on what is called the
maximal Riemannian submanifold of the supergroup Gl($N_v|N_v$) \cite{OTV}.
This is a combination of compact (as for the usual Goldstone bosons) and
non-compact integration domain (for the Goldstone bosons of the ghost-ghost
block). The factorization of $U(x)$ is still as above,
but $\xi(x)$ is no longer traceless since the singlet field
does not decouple in the quenched approximation.  To leading order in the
$\epsilon$-expansion the partition function in a fixed topological sector is
\bea
\hspace{-0.2cm}{\cal Z}_\nu\! =\! \!
\int\!\! dU_0 d\xi \left({\rm Sdet}U_0\right)^\nu \!\exp\!\left\{\!
{\Sigma \over 2} V {\rm Str } M \!\!\left( U_0 + U_0^{-1}\right) \right.
\nonumber \\
\hspace{-0.2cm} +\! \left. \int\!\! d^4x  \!\left[
-\frac{1}{2} {\rm Str } \left(\partial_\mu \xi \partial_\mu \xi \right) +
{m_0^2 \over 6} \Xi^2 + {\alpha \over 6}
\partial_\mu \Xi\partial_\mu \Xi \right] \right\},\nonumber
\eea
with $\Xi (x) = {\rm Str} \xi (x)$ and  Str denotes the supertrace.


An alternative to the supersymmetric method is the
replica method \cite{DS}, which amounts to taking the $N_f \rightarrow 0$
limit of the full QCD result. Chiral perturbation theory done this way
is equivalent to that of the supersymmetric method, but the Feynman rules
are somewhat simpler. Its shortcoming in the present context
is that the integral over the zero
momentum modes in general can only be performed through series expansions
\cite{DS1}. As a first check, we have confirmed the calculation of the
leading correction
to the chiral condensate in the $\epsilon$-expansion \cite{D01}, now
using the supersymmetric method. Likewise, as a check on the results
presented below we have performed all calculations both ways, and thus
compared the resulting series expansions from the replica method with
the closed expressions obtained by the supersymmetric method. In all
cases we find complete agreement.

Here we present the first quenched results for correlation functions
in the $\epsilon$-expansion. We begin with
the scalar and pseudoscalar correlation functions
$s^a(x)\equiv \langle S^a(x) S^a(0) \rangle$ and
$p^a(x)\equiv \langle P^a(x) P^a(0) \rangle$
to ${\cal O}(\epsilon^2)$, which is the leading order contribution to
the space-time dependent terms. The ${\cal O}(1)$ contributions are constant.
In a sector with fixed topology we find for the space integrals
of flavor singlets
\bea
s^0(t) =  {\rm const} - {\Sigma^2
\over 2 F_\pi^2} \!
\left[ G(t) a_- - \Delta(t) {a_+ + a_- -  4 \over 2}  \right]
\nonumber \\
p^0(t) = {\rm const} + {\Sigma^2
\over 2 F_\pi^2}\! \left[ G(t) a_+ - \Delta(t) {a_+ + a_- + 4 \over 2}
\right] \nonumber
\eea
where $V = L^3T$ and 
\bea
a_+ &=& 4 {{\Sigma'}_{\nu } (\mu) \over \Sigma} + 4
+ 4 { \nu^2 \over \mu^2},\cr
a_- &=& - {4\over \mu} {\Sigma_{\nu } (\mu) \over \Sigma} +
4 { \nu^2 \over \mu^2} ~,\nonumber
\eea 
with the chiral condensate \cite{OTV},
\bea
\frac{\Sigma_{\nu}(\mu)}{\Sigma} \!=\! \mu(I_{\nu}(\mu)K_{\nu}(\mu)
+ I_{\nu+1}(\mu)K_{\nu-1}(\mu)) + \frac{\nu}{\mu} .\nonumber
\eea
The functions  $G(t)$ and  $\Delta(t)$ are, with $\tau = {t\over T}$,
\bea
\Delta (t) & = & T  h_1 (\tau)\nonumber\\
 G (t) & = & - {m_0^2\over3} T^3  h_2 (\tau)
 + {\alpha\over 3} T h_1(\tau)\nonumber\\
 h_1 (\tau) &= &{1 \over 2} \left[ ( \tau - {1 \over 2} )^2 -
{1\over 12} \right] \nonumber\\
h_2 (\tau) & = & {1 \over 24} \left[ \tau^2 ( 1 - \tau )^2 -
{1\over 30} \right]. \nonumber
\eea
The origin of the functions $h_{1,2}(\tau)$ has been discussed in detail
in the literature \cite{HL}. As a check on the above result, 
in the limit $\mu \rightarrow \infty$ we find
$$p^0(t) \rightarrow
+ {2\Sigma^2\over  F_\pi^2} \left[ G(t)- \Delta(t)  \right] ,$$
which coincides with the leading order result from the $p$-expansion.
This is simply tree level propagation of the flavor singlet.

We have also computed the flavored correlation functions. The calculation
is more involved because in the supersymmetric formulation it is necessary
to work with at least
$N_v=2$ \cite{TV} since the sources contain two flavors. All details will
be given in a forthcoming publication. 
Here we just show a plot of the pseudoscalar
correlation function $p^a(t), ~a=1,2,3$ 
in the $\epsilon$ and $p$-expansions, for different values
of the quark mass close to the range of validity of both
regimes\footnote{The constant terms have here been
included only to leading order in the $\epsilon$ expansion.}.
As in the unquenched case, there is an overlap region at $M_\pi L \sim 1$
where both expansions are good if $m \Sigma V \gg 1$.

The volume dependence of the above formulae provides an extra handle to
extract the low-energy constants $\Sigma, F_\pi, m_0^2$ and
$\alpha$ in this regime. This technique has already been successfully
applied to extract the
constant $\Sigma$ from the spectral density of the Dirac operator
and from the quark condensate at finite volume \cite{sigma}.

\begin{figure}[htb]
\epsfig{file=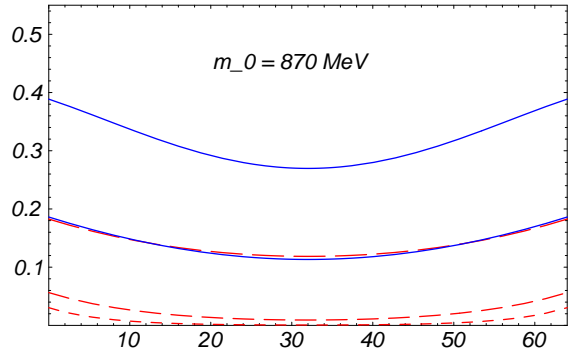,width=7.5cm}
\caption{Flavor-non-singlet $p^a(t)$ in lattice units with $a^{-1}
\sim 1.5 {\rm GeV}, V=32^3\times 64$. Shown is the $p$-expansion
(dashed lines, $m = 10, 3, 0.48 {\rm MeV}$) and the $\epsilon$-expansion (solid
lines, for $m = 0.48, 0.2 {\rm MeV}$). As $m$ is decreased the curves go
upward, as expected. In the intermediate mass region ($m \sim 0.5 {\rm Mev}$)
there is nice agreement between the two expansions.}
\label{fig}
\end{figure}

\end{document}